\documentclass[12pt]{article}

\usepackage{graphicx}
\usepackage{cite}

\begin{document}

%%-move to normal A4-%%
\hoffset = -1truecm \voffset = -2truecm \baselineskip = 8 mm

\title{\bf  Nucleon spin structure II: Spin structure function $g_1^p$ at small $x$}

\author{
 {\bf Wei Zhu} and {\bf Jianhong Ruan}\\
\\
\normalsize Department of Physics, East China Normal University,
Shanghai 200062, P.R. China \\}

\date{}

\newpage

\maketitle

\vskip 3truecm

\begin{abstract}

    The spin structure function $g_1^p$ of the proton
is studied in a two component framework, where the perturbative
evolution of parton distributions and nonperturbative vector meson
dominance model are used. We predict the $g_1^p$ asymmetric behavior at
small $x$ from lower $Q^2$ to higher
$Q^2$. We find that the contribution of the large gluon helicity
dominates $g_1^p$ at $x>10^{-3}$ but mixed with nonperturbative
component which complicates the asymptomatic behavior of $g_1^p$ at $x<10^{-3}$.
The results are compatible with the data including the HERA early estimations
and COMPASS new results.
The predicted strong $Q^2$- and
$x$-dependence of $g_1^p$ at $0.01<Q^2<3 GeV^2$
and $x<0.1$ can be checked on the next Electron-Ion Collider.

\end{abstract}

PACS number(s): 12.38.Cy, 12.38.Qk, 12.38.Lg, 12.40.Vv

$keywords$: Nucleon spin structure

\newpage
\begin{center}
\section{Introduction}
\end{center}

     Recently, COMPASS experiment at CERN collected a large number of events
of polarized inelastic scattering off the protons with very small
values of Bjorken scaling variable $x$ [1]. The preliminary analysis of these data combining with the previous experiments [2],
showed non zero and positive asymmetries of the structure function
$g_1^p$. In these fixed target experiments the low values of $x$ are
almost reached by lowering the values of $Q^2$. The knowledge of the
nucleon spin structure function $g_1(x,Q^2)$ at low $Q^2$ and small
$x$ is particulary interesting, since it is not only an important
information to resolve the "proton spin crisis", but also provides
us with a good place to study the transition from the perturbative
research to the nonperturbative description of the proton structure.

    In this work we try to study the behavior of $g_1^p$ at small $x$
but in the full $Q^2$ range. As we know that the structure functions
of the nucleon are mainly constructed of the parton distributions at
$Q^2>1 GeV^2$, while the non-perturbative contributions to the
structure functions become unneglectable at $Q^2\ll1~GeV^2$. A key
question is what components construct the spin structure functions of
the proton at such low $Q^2$? Particularly, do the parton
distributions and their pQCD evolution still play a role or not? For
answering these questions, we introduce the application of the
Dokshitzer-Gribov-Lipatov-Altarelli-Parisi (DGLAP) equation [3] with
the parton recombination corrections at low $Q^2$ in detail.
These corrections have been derived both for the polarized and
unpolarized parton distributions in our previous works [4,5]. We
point out that the isolation of the contributions of the vector
meson is necessary for keeping the factorization schema of the
polarized parton distributions at low $Q^2$. We find two different
asymptotic behaviors of $g_1^p$ at $x<10^{-3}$: nonperturbaive
behavior  $\sim x^{-1}$ at $Q^2<1 GeV^2$ and perturbative drop at
$Q^2>3 GeV^2$. We predict the translation of $g_1^p$ at small $x$
from lower $Q^2$ to higher $Q^2$. The results are compatible with
the data including the early HERA estimations and COMPASS new
results. We point out that the measurements at different $x$ with different
values of $Q^2$ in the fixed target experiments mix the complicated
asymptomatic behavior of $g_1^p$. The predicted strong $Q^2$- and
$x$-dependence of $g_1^p$ at $0.01<Q^2<3 GeV^2$
and $x<0.1$ due to the mixture of nonperturbative
vector meson interactions and the QCD evolution of the parton
distributions can be checked on the next Electron-Ion Collider (EIC).

     The organization of this paper is as follows. In Sec. 2
we discuss the applications of the polarized parton distributions at
low $Q^2$ based on the generalized leading order approximation. In
Sec.3 we summarize the contributions of the polarized parton
distributions to the spin structure function $g_1^p$ of the proton,
which have been fixed by our previous work. The contributions of the
vector meson to $g_1^p$ are discussed in Sec. 4. We present our
predictions of $g_1^p$ and the comparisons  with the data in
Sec. 5. The discussions and  summary are given in Sec. 6.

\newpage
\begin{center}
\section{A general consideration of the nucleon structure function at low $Q^2$ }
\end{center}

    In the researches of the nucleon structure functions at
the full kinematic region, an argued question is whether the parton
distributions and their perturbative QCD evolution can (even partly)
be applied to the low $Q^2$ range or the parton concept is suddenly
invalid at a critical value of $Q^2\leq 1GeV^2$?

    Let us begin from the parton model for the spin-dependent distribution, which
is written based on the Collins-Soper-Sterman (CSS) factorization
schema [6] at the collinear approximation and the twist-2 level,

$$g_1(x,Q^2)=\int^1_0\frac{dy}{y}\sum_qC_q(x/y,Q^2/\mu_F)\delta q(y,\mu_F),\eqno(2.1)$$
which breaks up the spin structure function into two factors
associated with perturbative short-distance functions $C_a$ and
nonperturbative polarized parton distributions $\delta q$ at the
factorization scale $\mu_F$.

 Taking the lowest order of $C_q$

$$C_q(x/y,Q^2/\mu_F)=\frac{1}{2}e_q^2\delta(x/y-1)\delta(Q-\mu_F)+{\cal O}(\alpha_s)+ {\cal O}(1/Q),\eqno(2.2)$$
${\cal O}(\alpha_s)$ and ${\cal O}(1/Q)$ are the QCD radiative
corrections and higher twist contributions. Inserting it to Eq. (2.1),
we obtain the relation between the spin structure functions and the
polarized quark distributions

$$g_1(x,Q^2)=\frac{1}{2}\sum_q e_q^2[\delta q(x,Q^2)+\delta \overline{q}(x,Q^2)]+{\cal O}(\alpha_s)+{\cal O}(1/Q). \eqno(2.3)$$
According to the renormalization group theory,

$$\frac{d g_1(x,Q^2)}{d\ln \mu_F}=0,\eqno(2.4)$$ it gives the DGLAP equation

$$Q^2\frac{d}{dQ^2}\delta q(x,Q^2)=\int^1_0\frac{dy}{y}\sum_{q'}\Delta P_{qq'}(x/y,\alpha_s(Q^2))\delta q'(y,Q^2),\eqno(2.5)$$
$\Delta P_{qq'}$ denotes the splitting functions. If we consider
only the leading order  $(LO$) approximation, we have

$$g_1^{DGLAP}(x,Q^2)=\frac{1}{2}\sum_q e_q^2[\delta q(x,Q^2)+\delta \overline{q}(x,Q^2)], \eqno(2.6)$$
These results are available at $Q^2>$ a few $GeV^2$.

    Now let us consider what will happen to the above results at low $Q^2$?
In principle, the initial parton distributions $\delta q(x,\mu^2)$ in
Eq. (2.1) are defined at $Q^2=\mu^2$ ($\Lambda_{QCD} <\mu<< 1GeV$).
Therefore, the parton distributions are
non-perturbative essentially. However, the factorization in Eq.(2.1) should be
modified at low $Q^2$ because of the following reasons:

    (i) The handbag diagram Fig.1a is a typical time ordered diagram
describing Eq. (2.1), where the quark propagators connecting with
the probe and the target have only the forward component, these
propagators can be broken as shown in Eq. (2.1) since they are
on-mass-shell. The corresponding backward quark propagators
construct the cat's ear diagram Fig.1b, which are neglected since these
backward propagators are absorbed by the target in the collinear
approximation [7]. However, the contributions of Fig. 1c to Eq.
(2.1) can not been neglected at low $Q^2$ due to the corrections of
quark-antiquark pair, which interacts with the target as a virtual vector
meson if the transverse momentum $k_\perp \sim Q$ of
quark pair is not large and confinement effects are essential. The
interference of the forward and backward quark propagators in Fig.
1a and 1c will put these propagators to off-mass-shell and breaks
the factorization schema. To avoid this event, we use a
phenomenological vector meson dominance (VMD) model [8] to "isolate"
the contributions from Fig. 1c. Traditionally, such VMD hypothesis
is used to explain the structure function at low $Q^2$ region
[9]. We denote this contribution as $g_1^{VMD}(x,Q^2)$.

    (ii) In CSS collinear factorization scheme the soft gluons
connect with the hard- and soft-parts can be absorbed into the
soft-part at the collinear approximation, where the transverse momentum
$k_T$ of the partons is neglected.  However, the $k_T$-effects of
the parton at low $Q^2$ should be considered. Thus, the collinear
factorization should be replaced by the $k_T$-factorization scheme.
Unfortunately, we haven't  a satisfy $k_T$-factorization scheme for
the spin structure functions. The $k_T$-effects also include the
replacements of $\delta q(x,Q^2)$ and the DGLAP equation with the
transverse momentum dependent (TMD) distribution $\delta
q(x,k_T,Q^2)$ and corresponding new evolution equations.
While we haven't such tools yet. We assume that a
satisfactory choice of the parameters in the input parton
distributions can mimic these $k_T$-effects.

    (iii) According to the operator product expansion
(OPE) in QCD [10], the $Q^2$-evolution of structure function moments
can be described in terms of twist expansion. The twist-2
represents the scattering from individual partons, while higher twist
corrections appear due to correlations among partons. At low $Q^2$
scale, the higher twist (HT) contributions to the structure functions
play a significant role. A special twist-4 corrections-the parton
recombination to the DGLAP evolution equation at $LL(Q^2)$
approximation has been derived by us for un-polarized and polarized
parton distributions in [4,5]. We will detail its contribution
$g_1^{DGLAP+ZRS}$ in next section. While the typical contributions
of the higher twist power suppressions $\sim
\sum_{i=2}^{\infty}\mu_{2i}(x,Q^2)/Q^{2i-2}$ to $g_1^p$ are
neglected in this work since they mainly change $g_1^p$ at $x>0.1$
[11].

    (iv) The more complicated corrections to $g_1$ at low $Q^2$ are from
the higher order QCD effects $\cal{O}(\alpha_s)$. In our works we
only consider the contributions of the leading order corrections. An
unavoidable question is whether we can neglect all higher order QCD
corrections when $Q^2\ll 1GeV^2$? Since all order resummation of these corrections
are difficult, we take following generalized
leading order (GLO) approximation [12]: if the leading order
contributions (or including necessary lowest order corrections) to a
given process are compatible with the experimental data, one can
conjecture that these neglected higher order corrections to this
process may cancelable each other, or they are successfully absorbed
by a finite number of free parameters.

    In consequence, at small $x$ and low $Q^2$ we have

$$g_1(x,Q^2)\simeq g_1^{DGLAP+ZRS}(x,Q^2)+g_1^{VMD}(x,Q^2),\eqno(2.7)$$
We will detail every term of Eq. (2.7) in next sections.

\newpage
\begin{center}
\section{Contributions of parton distributions}
\end{center}

    In this section, we present the contributions of the
polarized parton distributions of the proton to the spin structure functions at low $Q^2$, i.e.,

$$g_1^{DGLAP+ZRS}(x,Q^2)=\frac{1}{2}\sum_q e_q^2[\delta q(x,Q^2)+\delta
\overline{q}(x,Q^2)],\eqno(8)$$ where the QCD evolution dynamics
take the DGLAP equation with the parton recombination (ZRS)
corrections at the $LL(Q^2)$ approximation (see Eqs. (2.1)-(2.11)in
Ref.[12]), the minimum free parameters in the input parton
distributions have been fixed by the data mainly at $x>10^{-2}$,
they are

$$xu_v(x,\mu^2)=24.3x^{1.98}(1-x)^{2.06},\eqno(3.1)$$

$$xd_v(x,\mu^2)=9.10x^{1.31}(1-x)^{3.8},\eqno(3.2)$$ in [13], and

$$\delta u_v(x,\mu^2)=40.3x^{2.85}(1-x)^{2.15},\eqno(3.3)$$

$$\delta d_v(x,\mu^2)=-18.22x^{1.41}(1-x)^{4.0},\eqno(3.4)$$at
$\mu^2=0.064 GeV^2$ in [12]. The input distributions Eqs.
(3.1-3.4) neglect the contributions of asymmetry sea quark
corrections.

   Note that $1/\mu= 0.78 fm$ consists with a typical proton scale
$0.8-1~fm$. We consider that $\mu$ is a minimum transverse momentum
of the partons in the proton due to the uncertainty principle. Thus,
we assume that all parton distributions are freezed at scale $Q^2$
if $Q^2\le \mu^2$. Based on this assumption we avoid the un-physical
singularities at $Q\sim \Lambda_{QCD}$.

    We present $x$-dependence of $g_1^{DGLAP+ZRS}(x,Q^2)$ at
several values of $Q^2$ in Fig. 2. One can find the dramatic change
of the spin structure function at $x<10^{-3}$ from a flat form
to dramatically decreasing. Considering Fig. 3 in Ref.[12], we
conclude that the large gluon helicity effect leads to this
phenomenon.

\newpage
\begin{center}
\section{Contributions of the VMD part}
\end{center}

    As we have emphasized that the contribution
from the vector meson in virtual photon to $g_1^p$ at $Q^2<1GeV^2$
is necessary. Traditionally, this correction of the vector meson can
be described by the VMD model and it had been used for the predictions
of structure function at small $x$ and low $Q^2$ region.

    This contribution is written as

    $$xg_1^{VMD}(x,Q^2)=\frac{m_v^2}{8\pi}\sum_v\frac{m^2_vQ^2}{\gamma_v^2(Q^2+m^2_v)^2}\Delta\sigma_{vp}(s),\eqno(4.1)$$
where $\gamma_v$ is the coupling constant of vector meson and
proton; $x$ is a variable defined as $x=Q^2/(s+Q^2-m^2_p)$ rather
than a momentum fraction of parton, s is the CMS energy square for
the $\gamma p$ collision. The contributions of $\omega$ meson are
similar to that of $\rho$, while the contributions of $\phi$ meson
are small at $Q^2<1 ~GeV^2$ and can be neglected. We take $v=\rho$
and $\omega$ at $Q^2<1 GeV^2$ and $m_v=0.770 GeV$ . The
cross-sections $\Delta \sigma_{vp}(s)$ is the total cross section
for the scattering of polarized meson with the nucleon,
unfortunately, they are unknown. Usually, the following
parameterized formula is used,

$$\Delta\sigma_{vp}(s)\sim s^{\lambda-1}, at~x<x_0. \eqno(4.2)$$
Thus, we take

$$g_1^{VMD}(x,Q^2)=B\frac{(m^2_v)^{2-\lambda}(Q^2)^{\lambda}}{(Q^2+m^2_v)^2}[(\frac{x}{x_0})^{-\lambda}\theta
(x_0-x)+\frac{\ln^4x}{\ln^4x_0}\theta(x-x_0)],\eqno(4.3)$$ where the
second factor at $x>x_0$ is an arbitral function to suppress the
contributions of the VMD mechanism with increasing $x$. The
extrapolation of $g_1^p$ from the measured region down to $x\sim 0$
suggest us to assume that $\lambda=1-\epsilon$ and $x_0=10^{-3}$,
where $\epsilon\sim 0$ is a small positive parameter due to the
requirement of integrability of $g_1^p$ at $x\rightarrow 0$. In this
work, we temporarily take $\epsilon=0$. Thus,

$$g_1^{VMD}(x,Q^2)=B\frac{m^2_vQ^2}{(Q^2+m^2_v)^2}[(\frac{x_0}{x})\theta
(x_0-x)+\frac{\ln^4x}{\ln^4x_0}\theta(x-x_0)],\eqno(4.4)$$where the
parameter $B=3$.

\newpage
\begin{center}
\section{Predictions for spin structure function $g_1^p$ at small $x$}
\end{center}

     What is the asymptotic behavior of $g_1^p$? This is a broadly discussed
subject. We plot
$g_1^p(x,Q^2)=g_1^{DGLAP+ZRS}(x,Q^2)+g_1^{VMD}(x,Q^2)$ with
different values of $Q^2$ in Fig. 3. There are two different
asymptotic behaviors of $g_1^p$ at $x<10^{-3}$: the VMD behavior
$\sim x^{-1}$ at $Q^2<1 GeV^2$ and the large gluon helicity effect
at $Q^2>3 GeV^2$. Besides, $g_1^p$ presents the twist form of the
two asymptomatic behaviors above, which is the mixing result of the
nonperturbative and perturbative dynamics.

    We compare our predicted $g_1^p$ at $x>10^{-3}$ with the data [14] in
Fig. 4. These data on 2010 are more precise than the previous data. Note
that the values of $Q^2$ of every measured point are different and
they are taken from Table I of [14]. The theoretical curve is a smooth
connection among these points. This figure shows that the pQCD
evolution almost control the behavior of $g_1^p$ at $x>10^{-3}$.

    On the other hand, the combination of non-perturbative and
perturbative dynamics at $x<10^{-3}$ leads to a dramatic change of
$g_1^p$ around $Q^2=1\sim 3 GeV^2$.  Unfortunately, there are only several
data with large uncertainty about $g_1^p$ in this range. In
Figs. 5 and 6 we collect the HERA early data [15,16] at $Q^2=1$, $10
GeV^2$ which are un-generally used and compare them with our
predicted $g_1^p$. Figure 7 shows some of these data (trigon) [15]
and the comparisons with our results (dark points). Figure 8 is the
$Q^2$-dependence of $g_1^p$ with fixed $x$, the data are taken from
[17]. One can find that our predicted $g_1^p$ are compatible with
these data, although more precise measurements are necessary.

     Finally, we compare our results with the new COMAPSS (primary)
data [1,2] at $Q^2<1GeV^2$, which show that $g_1^p$ presents a
flat asymptomatic form at $x<10^{-3}$. This seems to contradict
with the predicted strong rise of $g_1^2$ at $Q^2<1GeV^2$ in Fig.3.
However, in the COMPASS fixed target experiments there is a strong
correlation between $x$ and $Q^2$, which makes low $x$ measurements
also with low $Q^2$. In Fig. 9 we take the average values of $Q^2$
for each probing values of $x$ (see Fig.1 in Ref.[1]). The results
are acceptable. Obviously, the measurements at different $x$ with
different values of $Q^2$ in the fixed target experiments mix two
different asymptomatic behaviors of $g_1^p$.

    We predict the stronger $Q^2$- and $x$-dependence of $g_1^p$ at $0.01<Q^2<3 GeV^2$
and $x<0.1$ due to the mixtion of nonperturbative
vector meson interactions and the QCD evolution of the parton
distributions in Fig. 3. For testing this prediction, the measurements of $g_1^p$ with fixed $x$ or
$Q^2$ at low $Q^2$ are necessary. The planning Electron-Ion Collider (EIC), for example,
eRHIC [18] and EIC@HIAF [19] can probe a broad low $Q^2<1GeV^2$-range, where we can check
the predicted behavior of $g_1^p$ at fixed $x$ or $Q^2$.

\newpage
\begin{center}
\section{Discussions}
\end{center}

    In general consideration, both the logs of $1/x$ and
$Q^2$ are equally important at small $x$ and low $Q^2$, and one should sum the double logarithmic
(DL) terms $(\alpha_s\ln^2(1/x))^n$, which predict the singular
behavior $g_1^p\sim x^{-\lambda}$ $(\lambda>0)$. To this end, some
special attentions are proposed [20]. For example, the double
logarithmic terms are taken into account via a suitable kernel of
the evolution equations in the infrared evolution equations, which was
first suggested by Lipatov [21], or alternatively taking a singular
initial parton distributions at $x<10^{-2}$, one can also mimic the
results of the DL-resummation.

     In this work, the behavior of $g_1^p$ at the same range is obtained
through a long evolution of the DGLAP equation with the parton
recombination corrections. We find that it is different from the
predictions of the DL-resummation, the asymptomatic behavior of the
polarized quark distributions at $x\rightarrow 0$ is controlled by
$\Delta P_{qg}$ in the DGLAP equation, rather than the
$\ln^k(1/x)$-corrections to the DGLAP-kernel. Thus, the difficult DL
resummation can be replaced by the fits of the initial quark
distributions $\delta q_v(x,\mu^2)$ in the DGLAP equation if the
evolution distance is long enough. This conclusion was also obtained
in the unpolarized structure functions [22].

       In summary, we use the DGLAP equation with the parton recombination corrections and
the nonperturbative vector meson dominance model to predict the spin
structure functions $g_1^p$ of the proton. We first present a
complete picture for the translation of $g_1^p$ from low $Q^2(\sim
0)$ to high $Q^2$ at small $x$.
We find that the contribution of the large gluon helicity
dominates $g_1^p$ at $x>10^{-3}$, but the mixtion with nonperturbative
component complicates the asymptomatic behavior of $g_1^p$ at $x<10^{-3}$.
The results are compatible with
the data including the early HERA estimations and COMPASS new
results. The predicted strong $Q^2$- and $x$-dependence of $g_1^p$ at $0.01<Q^2<3 GeV^2$
and $x<0.1$ due to the mixtion of nonperturbative
vector meson interactions and the QCD evolution of the parton
distributions can be checked on the next Electron-Ion Collider (EIC).

\newpage

\newpage

\begin{figure}[htp]

\vskip 3cm
\centering
\includegraphics[width=0.7\textwidth]{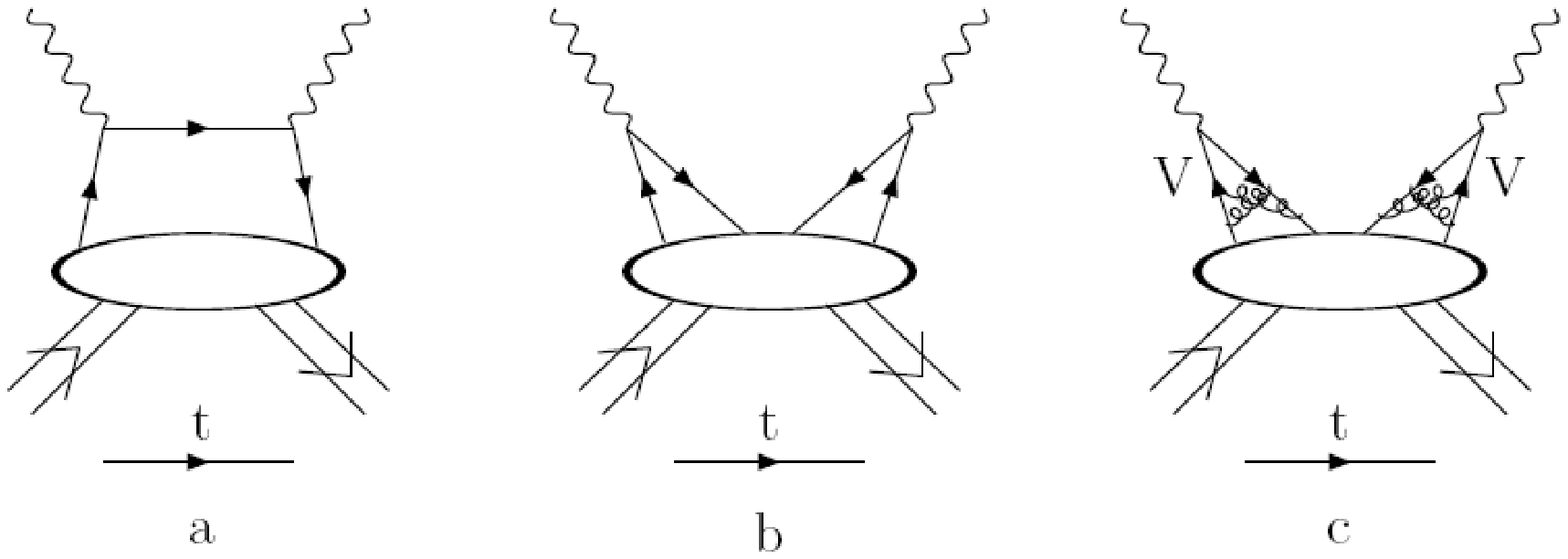}
\vskip 1cm
\caption{ The time ordered decomposing of DIS diagrams. (a) The struck
quarks are on-mass-shell since they have only forward component. (b)
A "cat ear" diagram, which vanishes in the collinear factorization
schema. (c) The "cat ear" diagram with higher order QCD corrections,
which are non-vanished at low $Q^2$, but can be isolated using a
naive VMD model.} \label{fig1}
\end{figure}

\begin{figure}[htp]
\centering
\includegraphics[width=0.7\textwidth]{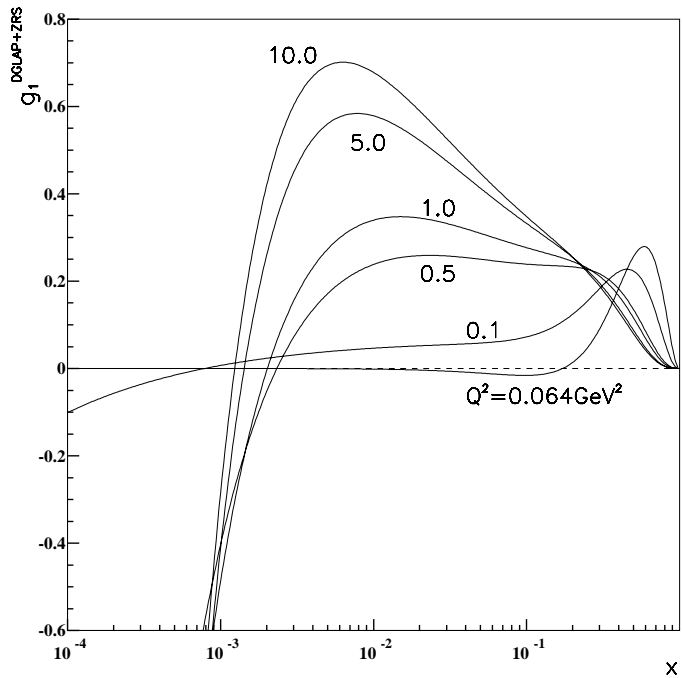}
\caption{Perturbative contributions $g^{DGLAP+ZRS}_1$ to $g_1^p$. All
partons are evolved from three valence quarks at $\mu^2=0.064GeV^2$.} \label{fig2}
\end{figure}

\newpage
\begin{figure}[htp]
\centering
\includegraphics[width=0.7\textwidth]{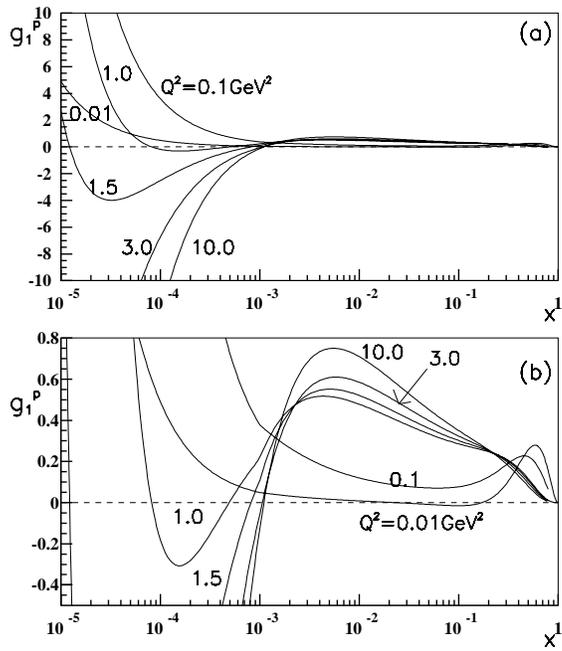}
\caption{$g_1^p$ evolutions at different values of $Q^2$ in (a) large
and (b) small scales.} \label{fig3}
\end{figure}

\newpage
\begin{figure}[htp]
\centering
\includegraphics[width=0.7\textwidth]{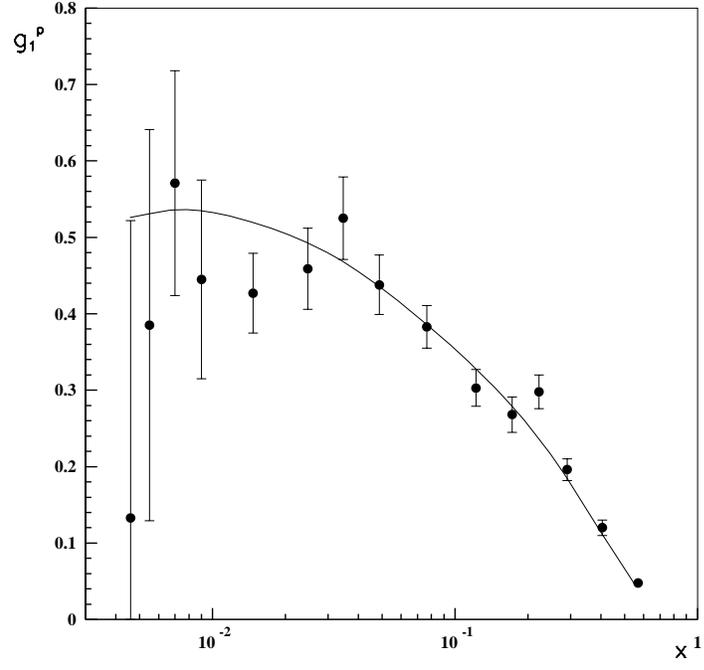}
\caption{Predicted $g^p_1$ at $x>10^{-3}$ and comparisons with the
COMPASS data [14]. Note that the values of $Q^2(x)$ of each measured
point are different (see Table I of Ref.[14]).}
\label{fig4}
\end{figure}

\newpage
\begin{figure}[htp]
\centering
\includegraphics[width=0.7\textwidth]{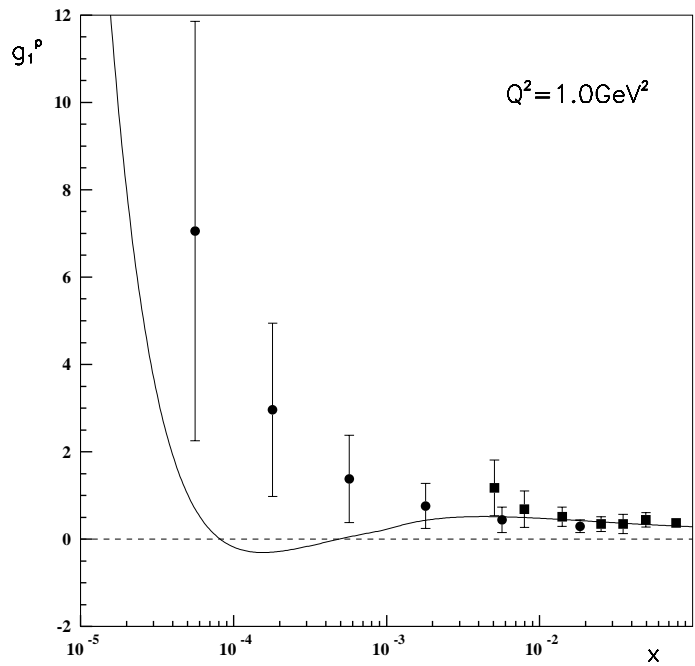}
\caption{Predicted $g_1^p$ at $Q^2=1GeV^2$ and the comparison with the
HERA data [15].}
\label{fig5}
\end{figure}

\newpage
\begin{figure}[htp]
\centering
\includegraphics[width=0.7\textwidth]{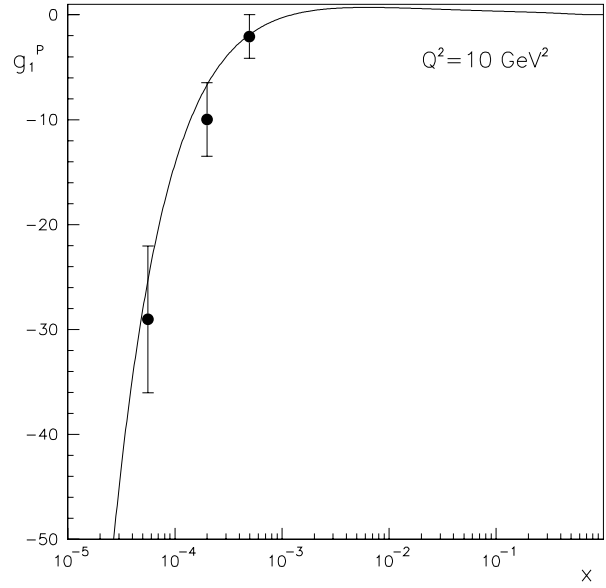}
\caption{Predicted $g_1^p$ at $Q^2=10GeV^2$ and the comparison with
the HERA "data", which are based on the NLO QCD predictions with the
statistical errors expected at HERA [16].} \label{fig6}
\end{figure}

\newpage
\begin{figure}[htp]
\centering
\includegraphics[width=0.7\textwidth]{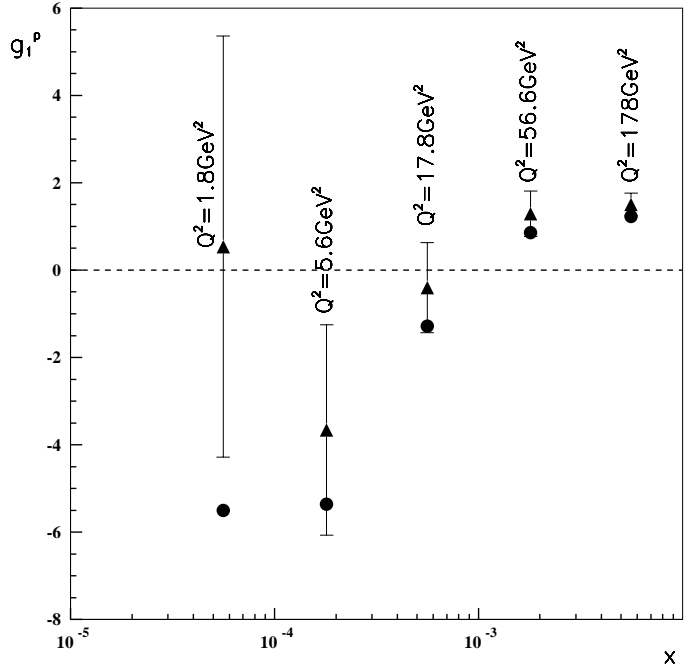}
\caption{Predicted $g_1^p$ at $Q^2=1.8GeV^2$, $5.6GeV^2$ and
$16.5GeV^2$ at $x<10^{-3}$ (circles) and the comparison with the
HERA data (triangles) [15].} \label{fig7}
\end{figure}

\newpage
\begin{figure}[htp]
\centering
\includegraphics[width=0.7\textwidth]{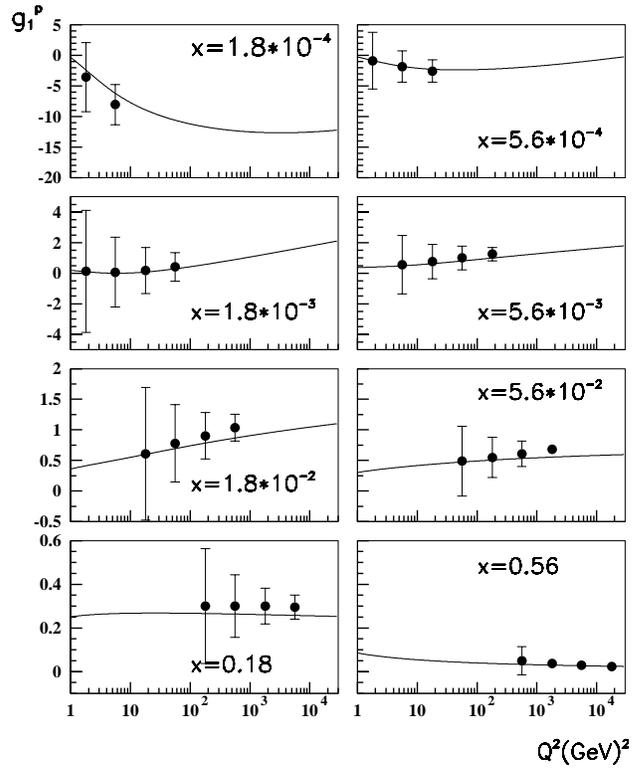}
\caption{Predicted $Q^2$-dependence of $g_1^p$ with fixed values of
$x$. Data are taken from [17]. } \label{fig8}
\end{figure}

\newpage
\begin{figure}[htp]
\centering
\includegraphics[width=0.7\textwidth]{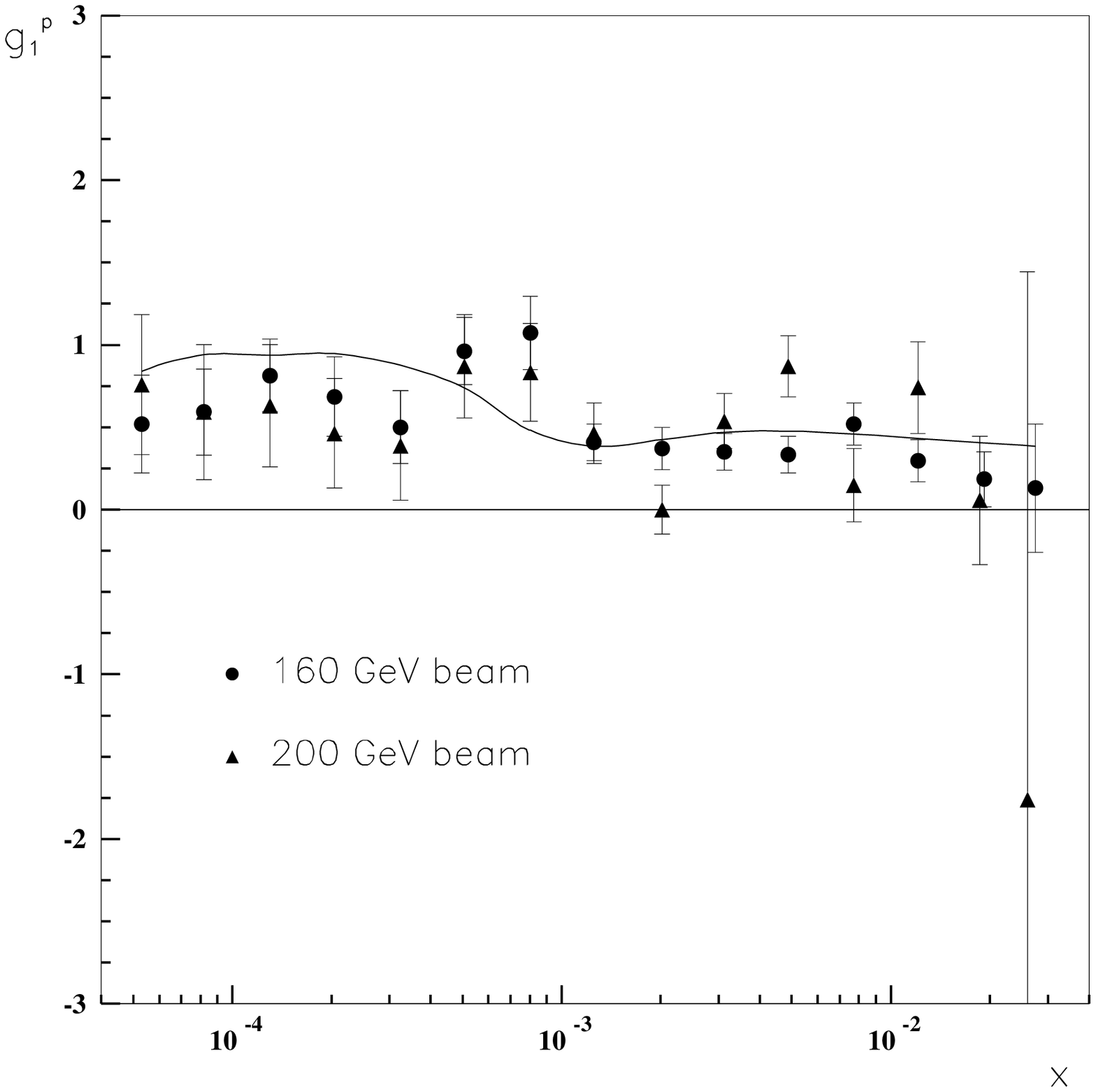}
\caption{Predicted $g^p_1$ as a function of $x$ with different
measured $Q^2(x)$ (solid curve). Note that the low values of $x$
connect with the low values of $Q^2(x)$. The data are
taken from COMPASS primary results with two different beam energies
[1]. } \label{fig9}
\end{figure}


\newpage
\begin{thebibliography}{99}

\bibitem{1} A.S. Nunes (on behalf of the COMPASS Collab.,)
Longitudinal double spin asymmetry $A^p_1$ and spin-dependent
structure function $g^p_1$ of the proton at low $x$ and low $Q^2$
from COMPASS,  Proceedings of the XV workshop on hihg energy spin
physics, Dubna, Russia, (2013), hep-ex/1405.5811.

\bibitem{2} COMPASS Collaboration, P. Abbon et al., Nucl. Instr. and Meth.
$\bf{A 577}$, 455 (2007); E.S. Ageev et al., Phys. Lett.
$\bf{B612}$, 154 (2005); V.Yu. Alexakhin et al., Phys. Lett.
$\bf{B647}$, 8 (2007); M.G. Alekseev et al., Phys. Lett.
$\bf{B690}$, 466 (2010).

\bibitem{3} G. Altarelli, G. Parisi, Nucl. Phys. $\bf{B126}$, 298 (1977);
V.N. Gribov, L.N. Lipatov, Sov. J. Nucl. Phys. $\bf{15}$, 438
(1972); Yu.L. Dokshitzer, Sov. Phys. JETP $\bf{46}$, 641 (1977).

\bibitem{4} W. Zhu, Nucl. Phys. $\bf{B551}$, 245 (1999), hep-ph/9809391;
W. Zhu, J.H. Ruan, Nucl. Phys. $\bf{B559}$, 378 (1999),
hep-ph/9907330v2; W. Zhu and Z.Q. Shen, HEP $\&$ NP, $\bf{29}$, 109
(2005), hep-ph/0406213v3.

\bibitem{5} W. Zhu, Z.Q. Shen and J.H. Ruan, Nucl. Phys. $\bf{B692}$, 417 (2004), hep-ph/0406212v2.

\bibitem{6} J.C. Collins, D.E. Soper, G. Sterman, in: A.H. Mueller (Ed.),
Perturbative Quantum Chromodynamics, World Scientific, Singapore,
1989, p. 1.

\bibitem{7} W. Zhu, H.W. Xiong, J.H. Ruan, Phys. Rev. $\bf{D60}$, 094006
(1999); W. Zhu, Nucl. Phys. $\bf{A753}$, 206 (2005).

\bibitem{8} J.J. Sakurai, currents and mesons, university of Chigag, Chigago (1969); T. H. Bauer et al.,
Rev. Mod. Phys. $\bf{50}$, 261 (1978); G. Grammer Jr and J. D.
Sullivan, in Electromagnetic Interactions of Hadrons, edited by A.
Donnachie and G. Shaw, Plenum, New York, 1978, Vol.2.

\bibitem{9} B. Badelek, J, Kwieci¨½ski, B. Ziaja, Eur. Phys. J.
$\bf{C26}$, 45 (2002); Acta Phys. Polon. $\bf{B33}$, 3701 (2002).

\bibitem{10} G. Sterman, An Introduction to Quantum Field Theory,
Cambridge Univ. Press, Cambridge, 1993.

\bibitem{11} E. Leader, A. V. Sidorov, D. B.
Stamenov, the proceedings of First Workshop on Quark-Hadron Duality
and the Transition to pQCD, Frascati, June 6-8, 2005,
hep-ph/0509183.

\bibitem{12}  W. Zhu and J.H. Ruan, Nucleon spin structure I:
a dynamical determination of polarized gluon distribution in the
proton, this serious works.


\bibitem{13}  X.R. Chen, J.H. Ruan, R. Wang, P.M. Zhang and W. Zhu, Int.
J. Mod. Phys. $\bf{E23}$, 1450057 (2014), hep-ph/1306.1872; ibit
$\bf{E23}$, 1450058 (2014), hep-ph/1306.1874.

\bibitem{14} COMPASS Collaboration, M.G. Alekseev, et al
Phys. Lett. $\bf{B690}$, 466 (2010).


\bibitem{15} R. D. Ball, A. Deshpande, S. Forte, V. W. Hughes, J. Lichtenstadt,, G.
Ridolf, Measurement of the polarized sreucture function
$g_1^p(x,Q^2)$ at HERA,  hep-ph/9609515.

\bibitem{16} J. Kwiecinski and B. Ziaja, hep-ph/9802386.

\bibitem{17} A. De Roeck, A. Deshpande, V.W. Hughes,
J. Lichtenstadt, G. Radel Eur. Phys. J. $\bf{C6}$, 121 (1999).

\bibitem{18} E.C. Aschenauer, at. al., eRHIC Design Study: An Electron-Ion Collider at BNL, hep-ph/1409.1633.

\bibitem{19} X.R. Chen, An Electro Ion Collider Plan in China, Invited talk at the
21st International Symposium on Spin Physics, Beijing, China, Oct. 20-24 (2014).

\bibitem{20} D. Kotlorz and A. Kotlorz,  Acta Phys. Polon. $\bf{B39}$, 1913 (2008);
B.I. Ermolaev, M. Greco, S.I. Troyan, Riv. Nuovo Cim. $\bf{33}$, 57
(2010).

\bibitem{21} L.N. Lipatov. Zh.Eksp.Teor.Fiz. $\bf{82}$, 991 (1982); Phys. Lett.
$\bf{B116}$, 411 (1982).

\bibitem{22} M. Gl$\ddot{u}$ck, E. Reya, and A. Vogt, Eur. Phys.
J. $\bf{C5}$, 461 (1998).





\end{thebibliography}
\end{document}